\def\etal{{\it et al.}}

\def\~{{$\tilde{\phantom{a}}$}}

\documentclass [12pt] {article}
\usepackage{epsfig}
\usepackage{color}

\def\thebibliography#1{\section{References}\markboth
 {REFERENCES}{REFERENCES}\list
 {[\arabic{enumi}]}{\settowidth\labelwidth{[#1]}\leftmargin\labelwidth
 \advance\leftmargin\labelsep
 \usecounter{enumi}}
 \def\newblock{\hskip .11em plus .33em minus -.07em}
 \sloppy
 \sfcode`\.=1000\relax}
\def\upcite#1{\raise6pt\hbox{\scriptsize
\cite{#1}}}
  \def\lsim{\mathrel {\vcenter {\baselineskip 0pt \kern 0pt
    \hbox{$<$} \kern 0pt \hbox{$\sim$} }}}
    \def\gsim{\mathrel {\vcenter {\baselineskip 0pt \kern 0pt
    \hbox{$>$} \kern 0pt \hbox{$\sim$} }}}

\def\hline{\noalign{\hrule \vskip2pt}}

\def\|{\ifmmode\Vert\else \char`\|\fi}
\ifx\oldzeta\undefined                          
  \let\oldzeta=\zeta                            
  \def\zzeta{{\raise 2pt\hbox{$\oldzeta$}}}     
  \let\zeta=\zzeta                              
\fi

\ifx\oldchi\undefined                           
  \let\oldchi=\chi                              
  \def\cchi{{\raise 2pt\hbox{$\oldchi$}}}       
  \let\chi=\cchi                                
\fi

\def\frac#1#2{{#1 \over #2}}

\def\half{\ifinner {\scriptstyle {1 \over 2}}
   \else {1 \over 2} \fi}

\def\simge{\mathrel{%
   \rlap{\raise 0.511ex \hbox{$>$}}{\lower 0.511ex \hbox{$\sim$}}}}
\def\simle{\mathrel{
   \rlap{\raise 0.511ex \hbox{$<$}}{\lower 0.511ex \hbox{$\sim$}}}}

\def\buildchar#1#2#3{{\null\!                   
   \mathop#1\limits^{#2}_{#3}                   
   \!\null}}                                    
\def\overcirc#1{\buildchar{#1}{\circ}{}}

\def\slashchar#1{\setbox0=\hbox{$#1$}           
   \dimen0=\wd0                                 
   \setbox1=\hbox{/} \dimen1=\wd1               
   \ifdim\dimen0>\dimen1                        
      \rlap{\hbox to \dimen0{\hfil/\hfil}}      
      #1                                        
   \else                                        
      \rlap{\hbox to \dimen1{\hfil$#1$\hfil}}   
      /                                         
   \fi}                                         %

\def\subrightarrow#1{
  \setbox0=\hbox{
    $\displaystyle\mathop{}
    \limits_{#1}$}
  \dimen0=\wd0
  \advance \dimen0 by .5em
  \mathrel{
    \mathop{\hbox to \dimen0{\rightarrowfill}}
       \limits_{#1}}}                           

\def\overlay#1#2{\ifmmode%
\setbox0=\hbox{$#1$}%
\setbox1=\hbox to\wd0{\hss$#2$\hss}\else%
\setbox0=\hbox{#1}%
\setbox1=\hbox to\wd0{\hss#2\hss}\fi%
#1\hskip-\wd0\box1 }

\def\pmb#1{\leavevmode\setbox0=\hbox{#1}%
\kern-.02em\copy0\kern-\wd0
\kern.04em\copy0\kern-\wd0
\kern-.02em\raise.04em\box0 }

\def\vereq#1#2{\lower3pt\vbox{\baselineskip1.5pt \lineskip1.5pt
\ialign{$\m@th#1\hfill##\hfil$\crcr#2\crcr\sim\crcr}}}

\def\tensor#1{\protect\@ontopof{#1}{\leftrightarrow}{1.15}\mathord{\box2}}
\def\overstar#1{\protect\@ontopof{#1}{\ast}{1.15}\mathord{\box2}}
\def\overdots#1{\protect\@ontopof{#1}{\cdots}{1.0}\mathord{\box2}}
\def\overcirc#1{\protect\@ontopof{#1}{\circ}{1.2}\mathord{\box2}}
\def\loarrow#1{\protect\@ontopof{#1}{\leftarrow}{1.15}\mathord{\box2}}
\def\roarrow#1{\protect\@ontopof{#1}{\rightarrow}{1.15}\mathord{\box2}}

\def\@ontopof#1#2#3{%
{\mathchoice
{\@@ontopof{#1}{#2}{#3}\displaystyle\scriptstyle}%
{\@@ontopof{#1}{#2}{#3}\textstyle\scriptstyle}%
{\@@ontopof{#1}{#2}{#3}\scriptstyle\scriptscriptstyle}%
{\@@ontopof{#1}{#2}{#3}\scriptscriptstyle\scriptscriptstyle}%
}%
}

\def\@@ontopof#1#2#3#4#5{%
\setbox0=\hbox{$#4#1$}%
\setbox1=\hbox{$#5#2$}%
\setbox2=\hbox{}\ht2=\ht0 \dp2=\dp0 %
\ifdim\wd0>\wd1 %
\setbox1=\hbox to\wd0{\hss\box1\hss}%
\mathord{\rlap{\raise#3\ht0\box1}\box0}%
\else   %
\setbox1=\hbox to.9\wd1{\hss\box1\hss}%
\setbox0=\hbox to\wd1{\hss$#4\relax#1$\hss}%
\mathord{\rlap{\copy0}\raise#3\ht0\box1}%
\fi
}%

\def\lambdabar{\protect\@lambdabar}
\def\@lambdabar{%
\relax
\bgroup
\def\@tempa{\hbox{\raise.73\ht0
\hbox to0pt{\kern.25\wd0\vrule width.5\wd0
height.1pt depth.1pt\hss}\box0}}%
\mathchoice{\setbox0\hbox{$\displaystyle\lambda$}\@tempa}%
{\setbox0\hbox{$\textstyle\lambda$}\@tempa}%
{\setbox0\hbox{$\scriptstyle\lambda$}\@tempa}%
{\setbox0\hbox{$\scriptscriptstyle\lambda$}\@tempa}%
\egroup
}

\def\corresponds{{\lower.2ex\hbox{=}}{\rm\kern-.75em^\triangle}}
\def\succsim{\succ\kern-.9em_\sim\kern.3em}
\def\precsim{\prec\kern-1em_\sim\kern.3em}
\def\slantfrac#1#2{\kern1em^{#1}\kern-.3em/\kern-.1em_{#2}}

\def\({ \left( }
\def\){ \right) }
\def\b{\begin{equation}}
\def\e{\end{equation}}
\def\={\ =\ }

\def\+{\ +\ }
\def\-{\ -\ }

\begin{document}     

\begin{center}
{\bf\LARGE A Strategy for Accelerator-Based Neutrino
\\

\medskip

Physics in the USA}

\medskip

K.T.~McDonald
\\
{\sl Joseph Henry Laboratories, Princeton University, 
Princeton, NJ 08544 USA}

\medskip

(April 29, 2002)

\bigskip


\end{center}

We outline a strategy for next-generation neutrino physics experiments based
on beams from accelerators in North America.  This strategy is based on the
mounting evidence in favor of the large mixing angle solution to solar
neutrino problem \cite{SNO}, which implies that in addition to measurement of
$\sin^2 2\theta_{13}$ and of the sign of $\Delta m^2_{23}$, measurement of CP violation in the neutrino sector is a realizable goal provided
$\sin^2 2\theta_{13}$ is not too small

The strategy is to begin with a new detector, 20-30 kton of liquid argon,
designed to make best use of the NUMI beam currently under construction at FNAL.
Then, after new measurements have made the optimal path clearer, we anticipate
choosing among options for neutrino ``superbeam" upgrades at BNL and/or FNAL 
as well as for second new detector of 100-200 kton.

\section{Basis of the Strategy}

The strategy is based on eight considerations of neutrino physics and
neutrino beams:
\begin{itemize}
\item
Improved measurements of the neutrino mixing parameters $\Delta m^2_{23}$,
and $\sin^2 2\theta_{23}$, as well as new measurements of 
$\sin^2 2\theta_{13}$, are best accomplished with a detector located at
the first oscillation maximum of $\nu_2 \leftrightarrow \nu_3$, namely
$L[{\rm km}] = 1.24 E_\nu[{\rm GeV}] / \Delta m^2_{23}[{\rm eV}^2]
\approx 500 E_\nu[{\rm GeV}]$, supposing $\Delta m^2_{23} = 2.5 \times
10^{-3}$ eV$^2$.

\item
Measurements of CP violation are possible (presuming the LMA solution
to the solar neutrino problem holds) with roughly equal accuracy
at any maximum of the $\nu_2 \leftrightarrow \nu_3$
oscillation pattern \cite{marciano}, but best accuracy is obtained
at the lowest energy practicable.

\item
The sign of $\Delta m^2_{23}$ can be determined via matter effects
\cite{wolf,ms}, which grow with distance but are very
small for $L \lsim 1000$ km.

\item
If $\Delta m^2_{12}$ is at or above the upper limit of the presently
allowed value in the LMA solution, as should be clarified in a year
or two by KamLAND \cite{kamland}, then two scales of oscillation
will be discernible in very long baseline experiments \cite{shrock}.

\item
Accelerator-based neutrino beams from pion (and kaon and muon) decay
are dominantly $\nu_\mu$ (from positive mesons, and $\bar\nu_\mu$ from
negative mesons) with admixtures of $\nu_e$ and $\bar\nu_\mu$ at the
one percent level.  These backgrounds limit the sensitivity of
measurements of $\sin^2 2\theta_{13}$ and CP violation, which
rely on detection of $\nu_\mu \to \nu_e$ oscillations.  Furthermore,
if the beam energy is high enough that $\nu_\mu \to \nu_\tau$ 
oscillations can materialize as $\tau$ leptons, the semileptonic
decay $\tau \to e X$ causes undesirable backgrounds.

\item
Accelerator-based neutrino beams will have a broad energy distribution,
unless special efforts are made to reduce this, which leads to background
to the $\nu_\mu \to \nu_e$ signal from $\nu_\mu$ neutral-current
interactions, and $\nu_\tau$ charged-current interactions,
of higher than nominal energy.

\item
If we are confident about the value of $\Delta m^2_{23}$, it is 
therefore advantageous to narrow the energy spectrum of the beam,
which can be accomplished for low-energy neutrinos by use of an
off-axis neutrino beam \cite{e889,offaxis} that 
enhances the neutrino flux at an angle $\theta \approx 2^\circ /
E_\nu[{\rm GeV}]$ (due to the Jacobian peak in the two-body decay
kinematics of the pion), as shown in Fig.~\ref{oa3}.

\begin{figure}[htp]  
\begin{center}
\includegraphics*[width=3in]{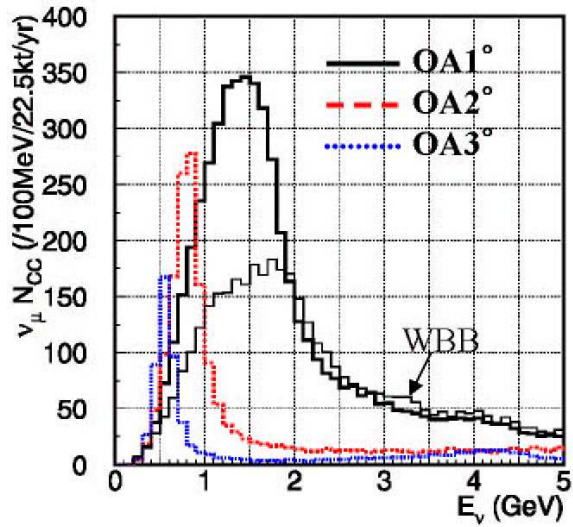}
\parbox{5.5in} 
{\caption[ Short caption for table of contents ]
{\label{oa3} Comparison of neutrino spectra for off axis beams at $1^\circ$,
$2^\circ$ and  $3^\circ$ with a wideband beam (WBB) derived from a
conventional neutrino horn \cite{JHF}.
}}
\end{center}
\end{figure}

\item
Interactions of neutrino of energies below about 700 MeV with nucleons
are primarily quasi-elastic with two body final states that permit
further suppression of neutral-current backgrounds \cite{JHF}.
A remaining troublesome background is inelastic scatters with a single
$\pi^0$ in the final state, whose decay photons can be mistaken for
electrons.

\end{itemize}

\section{Phase I of the Program}

In Phase I a 20-30 kton liquid argon detector is built at a site
near Soudan but $1^\circ$ off axis from the NUMI beam, with
emphasis on measurement of $\sin^2 2\theta_{13}$.

Based on the considerations of sec.~1, Phase-I of the
measurement strategy is formulated as follows:
\begin{itemize}

\item
The broadest program of measurement of neutrino oscillation
parameters in accelerator experiments should emphasize 
a low-energy
beam, $E_\nu < 1 $ GeV, and a detector at distance $< 400$ km,
while retaining the option for studies at longer baselines.

\item
However, to begin the program in a timely manner, it should start 
by using the NUMI beam from FNAL, whose spectrum drops rapidly
below 2 GeV, even when enhanced by going $\approx 1^\circ$ off axis. 

\item
Hence, the program should begin with a new detector, sited about $1^\circ$
off the nominal NUMI line, at about 700 km from FNAL so as to be near
the first oscillation associated with $\Delta m^2_{23}$.  A suitable site
would be near Silver Creek, MN (lat.\ 47.11$^\circ$, long.\ $-91.58^\circ$
\cite{topozone}, 640 km from FNAL), 
where a horizontal tunnel could be dug into a bluff next to Lake Superior to
provide $\approx 400'$ overburden.  See Fig.~\ref{bnl_sites}.

\begin{figure}[htp]  
\begin{center}
\includegraphics*[width=5in]{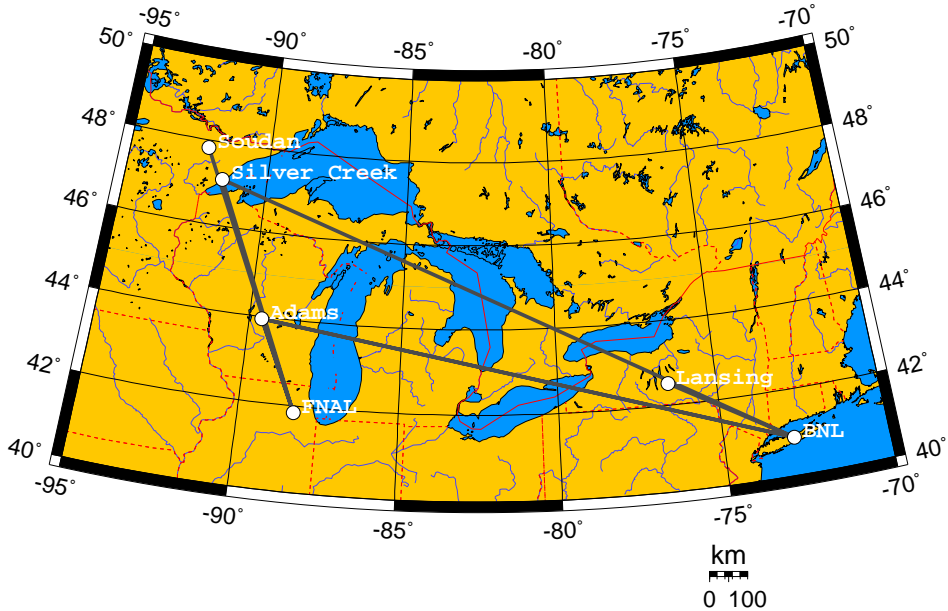}
\parbox{5.5in} 
{\caption[ Short caption for table of contents ]
{\label{bnl_sites} Possible sites for new detectors in neutrino beams
from FNAL and BNL.  In phase I a 20-30 kton liquid argon detector would
be built near Silver Creek, MN.  In Phase II a 100-200 kton detector 
would be built near Adams, WI or Lansing, NY and new neutrino beams
built at BNL and FNAL using 1-4 MW proton driver upgrades.
}}
\end{center}
\end{figure}

\item
Among various types of detectors, a liquid argon time projection
chamber \cite{Rubbia77}
has the best rejection of background due to neutral-current
interactions and low-energy $\pi^0$'s.
This permits measurements of $\sin^2 2\theta_{13}$ 
and CP violation via $\nu_\mu \to \nu_e$ oscillations with a much
smaller detector mass of argon, as illustrated in Fig.~\ref{harris}
from a recent study \cite{harris_nnn02}.

\begin{figure}[htp]  
\begin{center}
\includegraphics*[width=4in]{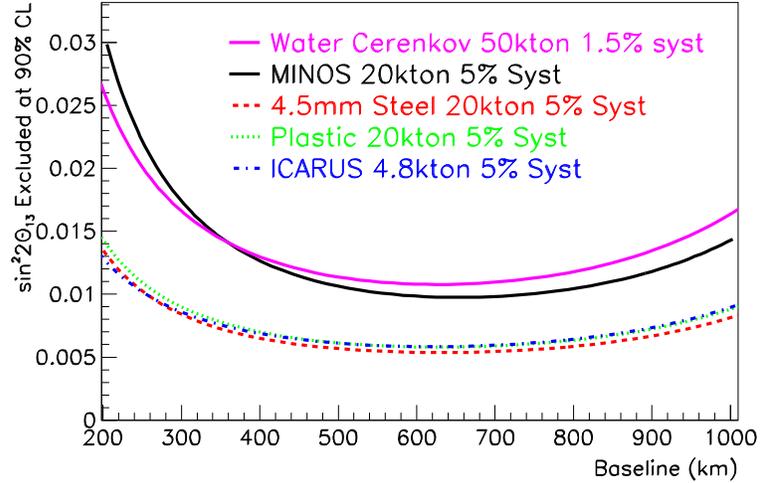}
\parbox{5.5in} 
{\caption[ Short caption for table of contents ]
{\label{harris} Comparison of several types of detectors in measuring
$\sin^2 2\theta_{13}$ in the presence of backgrounds typical of
a pion-decay neutrino beam at intermediate baselines \cite{harris_nnn02}.
The detector labeled ICARUS \cite{ICARUS}
is a liquid argon time projection chamber.  With a 25-kton liquid
argon detector, and an off-axis neutrino beam, 
as considered in the first phase of the present strategy,
the sensitivity to $\sin^2 2\theta_{13}$ would be at least 0.003.
}}
\end{center}
\end{figure}

\item
A 300-ton liquid argon detector has recently begun operation \cite{ICARUS},
with high-quality tracking of particle interactions as shown in
Fig.~\ref{icarus2}.
To extend measurements of neutrino oscillation parameters beyond those
that will be obtained by other experiments in the next decade, the 
initial detector mass must be at least 20 kton.  In later phases it will
be desirable to increase the detector mass to 100-200 kton, possibly at a
new site.   Cost are minimized when the detector is built as a single module,
such as that sketched in Fig.~\ref{lanndd1} \cite{lanndd,franco1}.

\begin{figure}[htp]  
\begin{center}
\includegraphics*[width=5in]{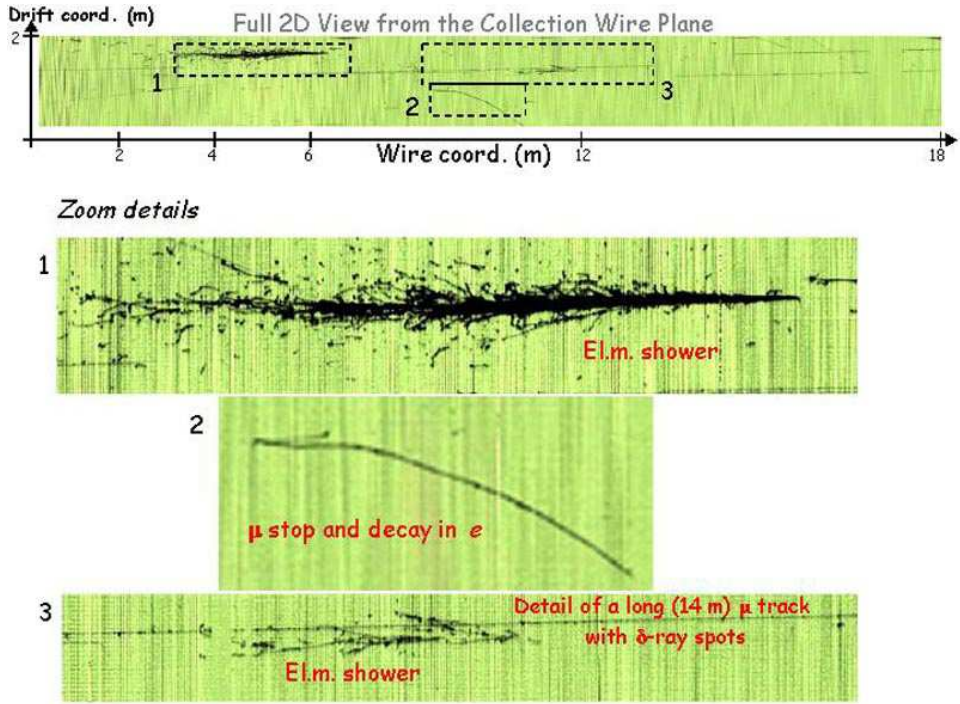}
\parbox{5.5in} 
{\caption[ Short caption for table of contents ]
{\label{icarus2} An event from the recent cosmic-ray test run of ICARUS  
\cite{ICARUS},
showing excellent track resolution over long drift distances in zero magnetic 
field.
}}
\end{center}
\end{figure}

\begin{figure}[htp]  
\begin{center}
\includegraphics*[width=4.5in]{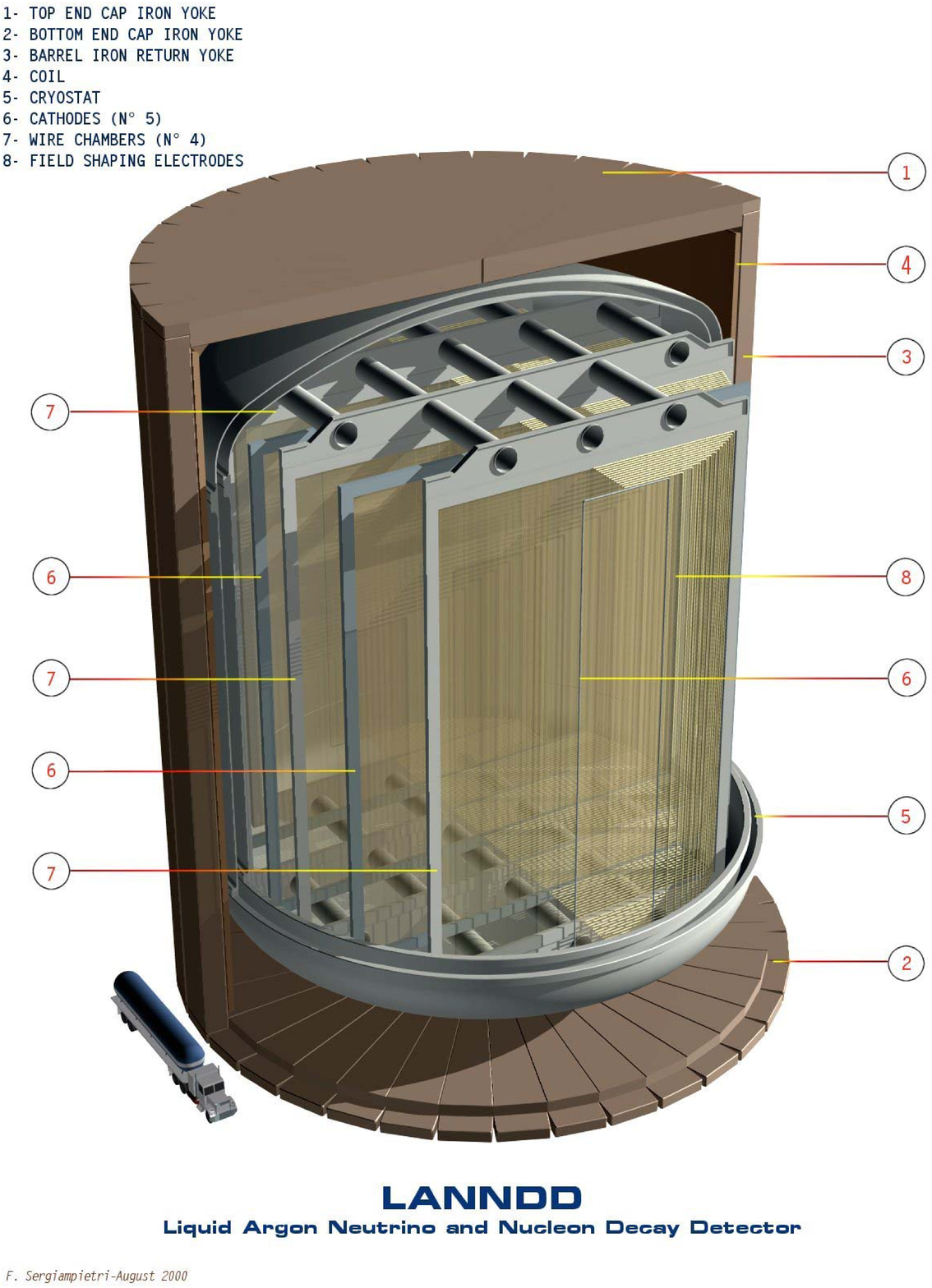}
\parbox{5.5in} 
{\caption[ Short caption for table of contents ]
{\label{lanndd1} Concept of a 70-kton Liquid Argon Neutrino and Nucleon Decay
Detector (LANNDD) \cite{lanndd,franco1}.
}}
\end{center}
\end{figure}

\item
The detector for an accelerator-based neutrino experiment could
be located on the surface of the Earth (or with a cover of $\approx
100$~m to suppress the cosmic-ray rate) because the beam duty factor
is $\approx 10^{-6}$.  Because of its excellent characterization of
events, even with only 100-m overburden a large liquid argon detector can provide essentially background-free sensitivity to proton decay via the modes
$p \to e^+ \pi^0$ and $p \to K^+ \bar\nu$ \cite{larpdk}.  
A 25 kton liquid argon detector yield a factor of 10
improvement in the sensitivity over present limits to proton decay 
via the mode $p \to K^+ \bar\nu$ \cite{pdk}
that is favored in generic SO(10) supersymmetric grand-unified models
\cite{Pati}.

\end{itemize}

\section{Phase II of the Program}

The first phase of the program, with a 20-30 kton liquid argon detector sited 
640 km from FNAL in a $1^\circ$ off-axis NUMI beam, will clarify whether the
next step is a) continued search for $\sin^2 2\theta_{13}$ as it is very small;
b) measurement of the sign of $\Delta M^2_{23}$; c) measurement of CP violation.
All three options will require increased neutrino flux and will benefit from
a larger detector.  Option 2 will require a longer baseline to enhance the
matter effects, while option 3 would benefit from a shorter baseline and 
lower energy beam.

A way of characterizing the goal of Phase II is 10 times the sensitivity to
$\sin^2 2\theta_{13}$ as in Phase I.  Since measurement of $\sin^2 2\theta_{13}$
is background limited, a factor of 10 improvement in sensitivity requires a
factor of 100 improvement in the product of neutrino flux and detector mass.
Thus, the general strategy is to upgrade the neutrino beam by a factor of 10 
(from a 0.4 MW proton driver to a 4 MW one), 
and also to increase the detector mass by 10 (from 20 to 200 kton).

There is considerable flexibility in the order of implementation of the
components of phase II.  The major choices are:
\begin{enumerate}
\item 
Build a new large detector close to FNAL, upgrade the FNAL beam to
operate at higher intensity and lower neutrino energy to study
$\sin^2 2\theta_{13}$ and CP violation, and build a neutrino beam at BNL
pointing to the new detector to study the sign of $\Delta M^2_{23}$.
\item
Build a new large detector close to BNL, build a new neutrino beam at
BNL to study $\sin^2 2\theta_{13}$ and CP violation in the new detector, 
and build an option into the BNL neutrino beam to be deflected slightly
to the Silver Creek detector to study the sign of $\Delta M^2_{23}$.
\end{enumerate}

If it appears appropriate to pursue measurement of the sign of $\Delta M^2_{23}$
before CP violation, the second scenario would be favored.

Scenario 1 is more costly than scenario 2 in that beam upgrades at both BNL and
FNAL are required.  The advantage of scenario 1 is more continuous operation of
the program, and the greater ultimate flexibility of having two high-performance
beams with different baselines to a single high-quality detector (plus
continued operation of the Phase-I detector at a 3rd baseline intermediate 
between the other two).  Scenario 2 has the ``cultural'' advantage that the
new detector could be located close to Cornell University, while the new
detector in scenario 1 will of necessity be at a somewhat remote location.

\subsection{Phase II with a New Detector near FNAL}

The ingredients of this scenario are:
\begin{enumerate}
\item
A 100-200 kton liquid argon detector sited closer to the neutrino source so
that lower-energy neutrinos can be used.  An example site is under a bluff
at lat.\ $43.954^\circ$, long.\ $-89.585^\circ$ near Adams, WI, as shown
in Fig.~\ref{bnl_sites} \cite{topozone}, where a
horizontal tunnel could be dug into the hillside to provide $\approx 100$ m
rock overburden.  This site is 260 km from FNAL, and $2.2^\circ$ off the NUMI
beam axis, which is appropriate for a neutrino beam of $\approx 700$ MeV.
\item
Upgrade of the FNAL proton driver to 4 MW.  As neutrinos of $\approx 700$ MeV
are desired, the proton beam energy need not be higher than 8 GeV.  Hence,
the needed proton driver upgrade is based on a high-performance 8-GeV booster
\cite{driver2}.
\item
Construction of a new neutrino beam at BNL with a proton driver of power 0.5-4
MW as the physics warrants \cite{bnl_loi,roser01}.  With this beam pointing
to the site of the new large detector, the sign of $\Delta M^2_{23}$ can be
measured, and additional data taken towards the measurement of CP violation.

\end{enumerate}

\subsection{Phase II with a New Detector near BNL}

The ingredients of this scenario are:
\begin{enumerate}
\item
The new 100-200 kton liquid argon detector could be sited in or near the
Cargill Salt Mine in Lansing, NY, lat.\ $42.509^\circ$, long.\ $-76.517^\circ$,
350 km from BNL,
as shown in Fig.~\ref{bnl_sites} \cite{Brierley}.   The working depth of
the mine is about $1700'$.  This site is in the
municipality immediately north of Cornell University. 

\item
Upgrade of the BNL proton driver to 4 MW, and construction of a new neutrino
beam \cite{bnl_loi,roser01}.  The beam would be pointed about $2^\circ$ below
the direction Lansing to obtain an enhancement of neutrino flux near 900 MeV.

\item
The angle between BNL-Lansing and BNL-Silver Creek is $5.9^\circ$ although the
azimuthal angle between Lansing and Silver Creek with respect to BNL
is only $1^\circ$.  Thus, neutrinos $4^\circ$ off-axis from the nominal beam
to Lansing would arrive at the Silver Creek detector.  This angle is slightly
too large to provide a useful neutrino flux.  Rather, when it is desired to
study the sign of $\Delta M^2_{23}$ with a beam from BNL to the Silver Creek
detector, the magnets of the target station would have to be rotated
downwards by about $2-3^\circ$.  That is, useful data could not be taken
simultaneously at the Lansing detector and the Silver Creek detector with a
beam from BNL.  Of course, the Silver Creek detector can continue to take
data with the NUMI beam from FNAL at all times.

\end{enumerate}

The conference version of this paper includes additional cartographical
information:
http://www.hep.princeton.edu/\~mcdonald/nufact/neutrinotrans12.pdf

\end{document}